\documentclass[10pt,conference]{IEEEtran}
\usepackage{amsmath,amssymb,pst-all}
\usepackage{epsfig,subfigure,color,graphicx}
\newtheorem{claim}{Claim }

\newtheorem{theorem}{Theorem}
\newtheorem{corollary}{Corollary}

\newcommand{\chsfk}[2]{\genfrac{[}{]}{0pt}{0}{#1}{#2}}

\renewcommand{\baselinestretch}{0.94}
\begin{document}

\title{On Wiretap Networks II}

\author{
\authorblockN{Salim Y. El Rouayheb}
\authorblockA{ECE Department \\
Texas A\&M University \\
College Station, TX 77843 \\
salim@ece.tamu.edu} \and
\authorblockN{Emina Soljanin}
\authorblockA{Math. Sciences Center \\
Bell Labs, Alcatel-Lucent \\
Murray Hill, NJ 07974 \\
emina@alcatel-lucent.com}
}

\maketitle

\begin{abstract}
We consider the problem of securing a multicast network against a
wiretapper that can intercept the packets on a limited number of arbitrary network links
of his choice.
We assume that the network implements network coding techniques
to simultaneously deliver all the packets
available at the source to all the destinations. We show how this problem can be looked at as a
network generalization of the Ozarow-Wyner Wiretap Channel of type II.
In particular, we show that network security can be achieved by using
the Ozarow-Wyner approach of coset coding at the source on top of the implemented
network code. This way, we quickly and transparently recover some of the
results available in the literature on secure network coding for wiretapped networks.
We also derive new bounds on the required secure code alphabet size and an algorithm
for code construction.
\end{abstract}

\section{Introduction}
Consider a communication network represented as a directed graph
$G=(V,E)$ with unit capacity edges, an information source $S$
that multicasts information to $t$ receivers $R_1,\dots,R_t$ located at distinct nodes.  Assume that
the min-cut value between the source and each receiver node is $n$. We know that a multicast rate of $n$ is
possible with linear network coding \cite{Ahl, Li}.
We are here concerned with  multicast networks in which there is an adversary that can
access data on a certain number of links of his choice, and the goal is to
maximize the multicast rate with the constraint of revealing no information
about the multicast data to the adversary.

The problem of making a linear network code information
theoretically secure in the presence of a wiretap adversary that can
look at a bounded number, say $\mu$, of network edges was first
studied by Cai and Yeung in \cite{yeung02secure}. They considered
directed graphs and demonstrated the existence of a code over an
alphabet with at least $\binom{|E|}{\mu}$ elements which can support
a secure multicast rate of up to $n - \mu$. They also showed that
such codes can be designed in ${\cal O}(\binom{|E|}{\mu})$ steps.
The required edge bandwidth and the secure code design complexity
are main drawbacks of this pioneering work. Feldman \emph{et al.}
derived trade-offs between security, code alphabet size, and
multicaat rate of secure linear network coding schemes in
\cite{feldman04csnc}, by using ideas from secret sharing and
abstracting network topology. Another approach was taken by Jain in
\cite{jain04} who obtained security by merely exploiting the
topology of the network in question. Weakly secure network coding
(which insures that only useless information rather than none is
revealed to the adversary) was studied by Bhattad and Narayanan in
\cite{bhattad05secure}, and practical schemes are missing in this
case as well.

A related line of work considers a more powerful adversary, one that can also modify the packets he observes.
Modifying a certain number of packets
in networks which only route information simply results in their incorrect reception, whereas
modifying the same number of packets carrying linear combinations of source packets can have a more harmful
effect since it can result in incorrect decoding of all source packets.
Such attacks are in network coding literature known as Byzantine modifications, and
the Byzantine modification detection in networks implementing random network coding was studied by
Ho \emph{et al.} in \cite{ho04byzantine} and Jaggi \emph{et al.} in \cite{jaggi07infocom}.
The approach they take is
to introduce error correction coding at the source so that the packets carry not only data but also
some redundant information derived from data which will help reduce the probability of incorrect decoding.

We also find coding at the source a natural approach to address the
information theoretic security of wiretap networks. In a network
where the min-cut value between the source and each receiver node is
$n$ and an adversary can access up to $\mu$ edges of his choice, we
introduce at the source a coding scheme which ensures information
theoretic security on the Ozarow-Wyner wiretap channel type II,
introduced in  \cite{Ozarow&Wyner:84} and \cite{OzarowWy85}, where
the source transmits $n$ symbols to the receiver and an adversary
can access any $\mu$ of those symbols.

Ozarow and Wyner showed that the maximum number of symbols (say $k$)
that the source can communicate to the receiver securely in the
information theoretic sense is equal to $n-\mu$. They also showed
how to encode the $k$ source symbols into the $n$ channel symbols
for secure transmission. Clearly, if the $n$ channel symbols are
multicast over a network not performing coding (linear combining of
the $n$ symbols), the $k$ source symbols remain secure in the
presence of an adversary with access to any $\mu$ edges. We will
illustrate later that this is is not necessarily the case when
network coding is performed. However, we will show that a network
code that preserves security of the $k$ source symbols (coded into
the $n$ multicast symbols in the Ozarow-Wyner manner) can be
designed over a sufficiently large field.

With the observations made by Feldman \emph{et al.}\ in
\cite{feldman04csnc}, it is easy to show that our scheme is actually
equivalent to the one proposed in the pioneering work of Cai and
Yeung in \cite{yeung02secure}. However, with our approach, we can
quickly and transparently recover some of the results available in
the literature on secure network coding for wiretapped networks.
Since the publication of \cite{yeung02secure} in which the network
code construction is based on the work of Li \emph{et al.}\ in \cite{Li}, a
number of simpler network code construction algorithms have been
proposed (see for example \cite{jaggi03polynomial}), \cite{ceitnc}. Computational
complexity of network coding in terms of the number of coding nodes
and ways to minimize it have also been studied since then
\cite{ceitnc}, \cite{lang06}, \cite{lang06com}. We will use these results to derive
new bounds on the required secure code alphabet size and an
algorithm for code construction.

This paper is organized as follows: In Sec.~\ref{sec:wtc}, we briefly review the Ozarow-Wyner wiretap channel type II problem.
In Sec.~\ref{sec:wtn}, we introduce the network generalization of this problem.
In Sec.~\ref{sec:cd}, we present an algorithm for secure network code design and discuss the required code alphabet size.
In Sec.~\ref{sec:con}, we highlight some connections of this work with the previous work on secure network coding and more recent work
on network error correction.

\section{Wiretap Channel II\label{sec:wtc}}
We first consider a point-to-point scenario in which the source can
transmit $n$ symbols to the receiver and an adversary can access any
$\mu$ of those symbols \cite{Ozarow&Wyner:84,OzarowWy85}. For this
case, we know that the maximum number of symbols that the source can
communicate to the receiver securely in the information theoretic
sense is equal to $n-\mu$.

The problem is mathematically formulated as follows. Let
$S=(s_1,s_2,\dots,s_k)$ be the random variable associated with the
$k$ information symbols that the source wishes to send securely,
$Y=(y_1,y_2,\dots,y_n)$ the random variable associated with the symbols
that are transmitted through the noiseless channel between the
source and the receiver, and $Z=(z_1,z_2,\dots,z_{\mu})$ the random
variable associated with the wiretapped bits of $Y$. When $k\le
n-\mu$, there exists an encoding scheme that maps $S$ into $Y$ so
that the uncertainty about $S$ is not reduced by the knowledge of
$Z$ and $S$ is completely determined (decodable) by the complete knowledge of $Y$, that is,
\begin{equation}
H(S|Z)=H(S) ~ \text{and} ~ H(S|Y)=0. \label{eq:seccon}
\end{equation}

For $n=2$, $k=1$, $\mu=1$, such a coding scheme can be organized as
follows. If the source bit equals $0$, then either $00$ or $11$ is
transmitted through the channel with equal probability. Similarly,
if the source bit equals $1$, then either $01$ or $10$ is
transmitted through the channel with equal probability.
\[
 \begin{array}{r cc }
 \text{source bit $s_1$:} &  0 & 1\\
\text{codeword $y_1y_2$ chosen at random from:} & \{00, 11\} & \{01,
10\}

\end{array}
 \]
It is easy to see that knowledge of either $y_1$ or $y_2$ does not
reduce the uncertainty about $s_1$, whereas the knowledge of both
$y_1$ and $y_2$ is sufficient to completely determine $s_1$, namely,
$s_1=y_1+y_2$.

In general, $k=n-\mu$ symbols can be transmitted securely by a
coding scheme based on an $[n,n-k]$ linear MDS code ${\cal
C}\subset\mathbb{F}_q^n$. In this scheme, the encoder is a
probabilistic device which operates on the space $\mathbb{F}_q^n$,
where $q$ is a large enough prime power,  partitioned into $q^k$
cosets of ${\cal C}$. The $k$ information symbols are taken as the
syndrome which specifies a coset, and the transmitted word is chosen
uniformly at random from the specified coset. The decoder recovers
the information symbols by simply computing the syndrome of the
received word. Because of the properties of MDS codes, knowledge any
 $\mu=n-k$ or fewer
symbols will leave uncertainty of the $k$ information symbols
unchanged. The code used in the above example is the $[2,1]$
repetition  with the parity check matrix
\begin{equation}
H=\begin{bmatrix}
    1 & 1 \\
  \end{bmatrix}.
\label{eq:hex}
\end{equation}

\section{Wiretap Network II\label{sec:wtn}}
We now consider again an acyclic multicast network $G=(V,E)$ with
unit capacity edges, an information source, $t$ receivers, and the
value of the mincut to each receiver equal to $n$. The goal is to
maximize the multicast rate with the constraint of revealing no
information about the multicast data to the adversary that can
access data on any $\mu$ links. We assume that the adversary knows
the implemented network code, \emph{i.e.}  all the coefficients of the
linear combinations that determine the packets on each edge.
Moreover, the adversary is aware of any shared randomness between the source
and the destinations. The last assumption rules out the use of
traditional "key" cryptography to achieve security.

 We know that a multicast rate of $n$
is possible with linear network coding \cite{Ahl, Li}.  It is
interesting to ask whether, using the same network code, the source
can multicast $k\le n-\mu$ symbols securely if it first applies a
secure wiretap channel code (as described above) mapping $k$ into
$n$ symbols. Naturally, this would be a solution if a multicast rate
of $n$ can be achieved just by routing.

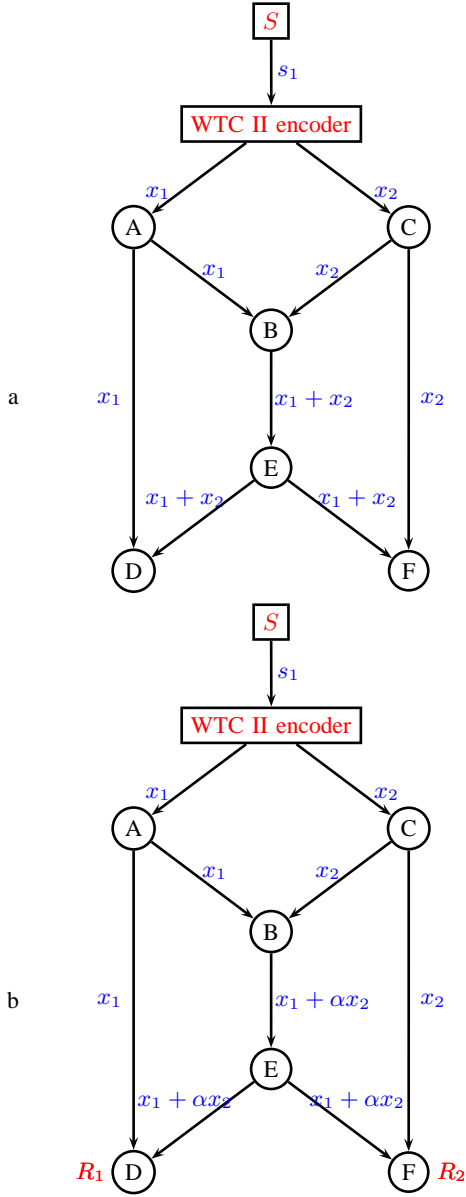
\begin{figure}[hbt]
\begin{center}
\psset{unit=0.090in}
\begin{pspicture}(0,5)(30,78)
\psset{linewidth=1.0pt}
\begin{small}
\rput(16,39){\rnode{S}{\psframebox{\textcolor{red}{$S$}}}}
\rput(16,33){\rnode{SE}{\psframebox{\textcolor{red}{WTC II
encoder}}}} \rput(8,27){\circlenode{A}{A}}
\rput(16,21){\circlenode{B}{B}} \rput(24,27){\circlenode{C}{C}}
\rput(8,7){\circlenode{D}{D}} \rput(16,13){\circlenode{E}{E}}
\rput(24,7){\circlenode{F}{F}} \rput(5.5,7){\textcolor{red}{$R_1$}}
\rput(26.5,7){\textcolor{red}{$R_2$}}
\ncline{->}{S}{SE} \ncline{->}{SE}{A} \ncline{->}{SE}{C}
\ncline{->}{A}{B}\ncline{->}{C}{B}
\ncline{->}{A}{D}\ncline{->}{C}{F}
\ncline{->}{E}{D}\ncline{->}{E}{F} \ncline{->}{B}{E}
\rput(17,71){\textcolor{blue}{$s_1$}}
\rput(9.4,64){\textcolor{blue}{$x_1$}}
\rput(22.7,64){\textcolor{blue}{$x_2$}}
\rput(12.7,59.4){\textcolor{blue}{$x_1$}}
\rput(19.3,59.4){\textcolor{blue}{$x_2$}}
\rput(6.6,52){\textcolor{blue}{$x_1$}}
\rput(25.4,52){\textcolor{blue}{$x_2$}}
\rput(18.5,52){\textcolor{blue}{$x_1+x_2$}}
\rput(11,46.2){\textcolor{blue}{$x_1+ x_2$}}
\rput(21,46.2){\textcolor{blue}{$x_1+x_2$}}

\rput(16,74){\rnode{S}{\psframebox{\textcolor{red}{$S$}}}}
\rput(16,68){\rnode{SE}{\psframebox{\textcolor{red}{WTC II
encoder}}}} \rput(8,62){\circlenode{A}{A}}
\rput(16,56){\circlenode{B}{B}} \rput(24,62){\circlenode{C}{C}}
\rput(8,42){\circlenode{D}{D}} \rput(16,48){\circlenode{E}{E}}
\rput(24,42){\circlenode{F}{F}} \rput(5.5,7){\textcolor{red}{$R_1$}}
\rput(26.5,7){\textcolor{red}{$R_2$}}
\ncline{->}{S}{SE} \ncline{->}{SE}{A} \ncline{->}{SE}{C}
\ncline{->}{A}{B}\ncline{->}{C}{B}
\ncline{->}{A}{D}\ncline{->}{C}{F}
\ncline{->}{E}{D}\ncline{->}{E}{F} \ncline{->}{B}{E}
\rput(17,36){\textcolor{blue}{$s_1$}}
\rput(9.4,29){\textcolor{blue}{$x_1$}}
\rput(22.7,29){\textcolor{blue}{$x_2$}}
\rput(12.7,24.4){\textcolor{blue}{$x_1$}}
\rput(19.3,24.4){\textcolor{blue}{$x_2$}}
\rput(6.6,17){\textcolor{blue}{$x_1$}}
\rput(25.4,17){\textcolor{blue}{$x_2$}}
\rput(19,17){\textcolor{blue}{$x_1+\alpha x_2$}}
\rput(11,11.2){\textcolor{blue}{$x_1+ \alpha x_2$}}
\rput(21,11.2){\textcolor{blue}{$x_1+\alpha x_2$}}
\rput(1,52){a}
\rput(1,17){b}
\end{small}
\end{pspicture}
\end{center}
\caption{Single-edge wiretap butterfly network
with a) insecure network code and b) secure network code.} \label{fig:butterfly_sec}
\end{figure}

Consider this approach for the butterfly network
shown in Fig.~\ref{fig:butterfly_sec}
where we have $n=2$, $k=1$, $\mu=1$. If the source applies the
coding scheme described in the previous section and the usual
network code as in Fig.~\ref{fig:butterfly_sec}-a, the adversary
will be able to immediately learn the source bit if he taps into any
of the edges BE, EF, ED. Therefore, a network code can brake
down a secure wiretap channel code. However, if the network code is
changed so that node B combines its inputs over {\it e.g.,}
$\mathbb{F}_3$ and the BE coding vector is $\begin{bmatrix}
    1 & \alpha \\
  \end{bmatrix}
$ where $\alpha$ is a primitive element of $\mathbb{F}_3$ (as in Fig.~\ref{fig:butterfly_sec}-b),
the wiretap channel code remains secure, that is, the adversary cannot gain any information by accessing
any single link in the network. Note that the wiretap channel code based on the MDS code with
$H=\begin{bmatrix}
    1 & 1 \\
  \end{bmatrix}$
remains secure with any network code whose BE coding vector is linearly
independent of $\begin{bmatrix}
    1 & 1 \\
  \end{bmatrix}
  $.

We will next show that the source can multicast $k\le n-\mu$ symbols
securely if it first applies a secure wiretap channel code based on
an MDS code with a $k\times n$ parity check matrix $H$ if the
network code is such that no linear combination of $\mu=n-k$ or
fewer coding vectors belongs to the space spanned by the rows of
$H$. Let $W\subset E$ denote the set of $|W|=\mu$ edges the
wiretapper chooses to observe, and $Z_W=(z_1,z_2,\dots,z_{\mu})$ the
random variable associated with the packets carried by the edges in
$W$. Let $C_W$ denote the matrix whose rows are the coding vectors
associated with the observed edges in $W$. As in the case of wiretap
channel, $S=(s_1,s_2,\dots,s_k)$ denotes the random variable
associated with the $k$ information symbols that the source wishes
to send securely, and $Y=(y_1,y_2,\dots,y_n)$ the random variable
associated the $n$ wiretap channel code symbols. The $n$ symbols of $Y$
will be multicast through the network by using linear network coding. Consider
$H(S,Y,Z_W)$ with the security requirement $H(S|Z_W)=H(S)$
for all $W\subset E$:
\begin{align*}
&\underbrace{H(S|Z_W)}_{=H(S)}+H(Y|SZ_W)=H(Y|Z_W)+\underbrace{H(S|YZ_W)}_{=0}\\
&\Rightarrow
H(Y|SZ_W) = H(Y|Z_W) - H(S)\\
&\Rightarrow
0\le n-\text{rank}(\mathbf{C}_W)-k
\end{align*}
Since there is a choice of edges such that $\text{rank}(C_W)=\mu$, the maximum rate
for secure transmission is bounded as
\[
k\le n-\mu.
\]
If the bound is achieved with equality, we have $H(Y|SZ_W)=0$ and consequently, the system of equations
\[
\begin{bmatrix}
  S \\
  Z_w \\
\end{bmatrix}
= \begin{bmatrix}
    H \\
    C_W \\
  \end{bmatrix}
  \cdot Y
\]
has to have unique solution for all $W$ for which $\text{rank}(C_W)=\mu$. That is,
\begin{equation}
\text{rank}
\begin{bmatrix}
    H \\
    C_W \\
  \end{bmatrix}
  = n ~~ \text{for all
  $C_W$ s.t.} ~ \text{rank}(C_W)=\mu.
\label{eq:secc}
\end{equation}
This analysis essentially proves the following result:
\begin{theorem}
\label{th:oursec} Let $G=(V,E)$ be an acyclic multicast network with
unit capacity edges, an information source and the  mincut value to
each receiver equal to $n$. A wiretap code at the source based on an
MDS code with a $k\times n$ parity check matrix $H$ and a network
code such that no linear combination of $\mu=n-k$ or fewer coding
vectors belongs to the space spanned by the rows of $H$ make the
network information theoretically secure against a wiretap adversary
who can observe at most $\mu \le n-k$ edges. Any adversary able to observe
more than $n-k$ edges will have uncertainty about the source smaller
than $k$.
\end{theorem}

The above analysis shows that the maximum throughput can be achieved
by applying a wiretap channel code at the source and then designing
the network code while respecting certain constraints. The decoding
of secure source symbols $S$ is then merely matrix multiplication of
the decoded multicast symbols $Y$. The
method gives us a better insight of how much information the
adversary gets if he can access more edges than the code is designed
for. It also gives us an insight on how to simply design secure
network codes in some cases over much smaller alphabets then
currently deemed necessary. Both claims are illustrated in the
example below.

\section{Network Code Design Alphabet Size\label{sec:cd}}
The approach described previously in the literature for finding
a secure multicast network code consisted of decoupling the problem
of designing a multicast network code and making it secure by using
some code on top of it.
Feldman \emph{et al.}\ showed in \cite{feldman04csnc}
that there exist networks where the above construction might require
a quite large field size. We investigate here a different
construction that, as was hinted in the conclusion of
\cite{feldman04csnc}, exploits the topology of the network. This is
accomplished  by incorporating the security constraints in the
\emph{Linear Information Flow} (LIF) algorithm of \cite{jaggi03polynomial} that
constructs linear multicast network codes in polynomial time in the number of edges
in the graph. The result is a better lower bound on the sufficient field size.
However, the modified LIF algorithm does not have polynomial time complexity.

We start by giving a brief high level overview of the LIF algorithm of
\cite{jaggi03polynomial}. The inputs of the algorithm are the network, the source node,
the $t$ destination nodes and the number $n$ of packets that need to be multicast to all
the destinations. Assuming the min-cut between the source and any
destination is at least $n$, the algorithm outputs a linear network
code that guaranties the delivery of the $n$ packets to all the
destinations.

The algorithm starts by 1) finding $t$ flows $F_1, F_2,\dots, F_t$
of value $n$ each, from the source to to each destination and 2)
setting $t$ $n \times n$ matrices $B_{F_j}$ (one for each receiver)
equal to $I_{n\times n}$ Then, it goes over the network edges,
visiting each one in  topological order.  In each iteration, the
algorithm finds a suitable local encoding vector for the visited
edge, and updates the  $t$ matrices $B_{F_j}$, each formed by the
global encoding vectors of the $n$ last visited edges in the flow
$F_j$.  The algorithm maintains the invariant that the matrices
$B_{F_j}$ remain invertible after each iteration. Thus, when it
terminates, each destination will get $n$ linear combination of the
original packets that form a full rank system. Thus each destination
can solve for these packets by inverting the corresponding system.

An important result of the previous algorithm, is that a field of
size at least  $t$ (the number of destinations) is always sufficient
for finding the desired network code. As  shown in \cite[Lemma
8]{jaggi03polynomial}, this follows from the fact that a field of
size larger or equal to $t$ is actually sufficient for satisfying
the condition that the $t$ matrices $B_{F_j}$ are always
invertible.

We modify the LIF algorithm so it  outputs a secure network code in
the following way. We fix the $k\times n$ parity check matrix $H$. WLOG, we assume that the $\mu$ packets observed
by the wiretapper are linearly independent, \emph{i.e.}\ rank $C_W=\mu$.
We denote by $e_i$ the edge visited at the $i$-th
iteration of the LIF algorithm, and by $P_i$ the set of the edges
that have been processed by the end of it. Then, we extend the set
of invariants to make sure that the encoding vectors are chosen so
the matrices $M_W=\chsfk{H}{C_w}$ are also invertible; which by
Theorem \ref{th:oursec} achieves the security condition. More
precisely, using the same techniques as the original LIF algorithm,
we make sure that by the end of the ith iteration, the matrices
$B_{F_j}$ and the matrices $M_{W_i}$ are invertible; where
$W_i=\{e_i\}\cup W'$ and $W'$ is a subset of  $P_i$ of order
$\mu-1=n-k-1$. The total number of the matrices that need to be kept
invertible in this modified version of the LIF algorithm is at most
$\binom{|E|-1}{\mu-1}+t$ (which corresponds to the last iteration).
Thus, similarly as in \cite[Lemma 8]{jaggi03polynomial}, we obtain
the following improved bound on the alphabet size for secure
multicast:
\begin{theorem}
Let $G=(V,E)$ be an acyclic network with
unit capacity edges, an information source, and the  mincut value to
each of the $t$ receivers equal to $n$.
A secure mulitcast at rate $k\le n$ in the presence of a wiretapper
who can observe at most $\mu\le n-k$ edges
is always possible over the alphabet $\mathbb{F}_q$ of size
\begin{equation}
q>\binom{|E|-1}{\mu-1}+t.
\label{eq:ourfs}
\end{equation}
\label{thbetterbound}
\end{theorem}

Bound (\ref{eq:ourfs}) can be further improved by realizing as was
first done in \cite{ceitnc} that not all edges in the network carry
different linear combination of source symbols. Langberg \emph{et
al.}\ showed in \cite[Thm.~5]{lang06} that the problem of finding
multicast network codes for a network $G$ can be reduced to solving
the same problem for a special equivalent network $\widehat{G}$ with
same parameters $n$ and $t$, which has the properties that all nodes
except the source and the destinations have total degree 3 and at
most $n^3t^2$ of its nodes have in-degree 2. These nodes are called
\emph{encoding nodes}, whereas the other ones are called
\emph{forwarding nodes} since the packets carried by their outgoing
edges are just copies of the ones available at their single incoming
edge. Given a network code for $\widehat{G}$, a one for $G$ can be
found efficiently over the same field. And, the set of global
encoding vectors of the edges of $G$ would be a subset of the one of
$\widehat{G}$.

Going back the security problem over a network $G$, one can try to
find a secure network code for the equivalent network $\widehat{G}$,
and then use the procedure described in \cite{lang06} and
\cite{lang06com} to construct a network code for $G$ which will also
be secure. Now consider the problem of finding secure network codes
for $\widehat{G}$. This problem will not change if the wiretapper is
not allowed to wiretap the \emph{forwarding edges}. Therefore, the
set of edges that the wiretapper might have access to consists of
the encoding edges and the edges outgoing from the source, and is of
order $n^3t^2+\delta$, where $\delta$ is the out-degree of the
source. Now, applying Theorem \ref{thbetterbound} on $\widehat{G}$
and taking into consideration the restriction on the edges that can
be potentially wiretapped, we obtain the following bound on the
sufficient field size which is independent of the size of the
network.
\begin{corollary}
For the transmission scenario of Thm.~\ref{thbetterbound}, a secure mulitcast network code always exists over the alphabet
$\mathbb{F}_q$ of size
\begin{equation}
q>\binom{k^3t^2+\delta}{\mu-1}+t. \label{eq:ourifs}
\end{equation}
\end{corollary}
\vspace{2mm}

For networks with two sources, we can completely settle the question on the required alphabet
size for a secure network code. Note that the adversary has to be limited to observing at most one edge of his
choice. Based on the work of Fragouli and Soljanin in \cite{ceitnc},
the coding problem for these networks is equivalent to a vertex coloring problem of some specially
designed graphs, where the colors are actually the points on
the projective line $\mathbb{PG}(1,q)$:
\begin{equation}
[0\, 1], ~ [1\, 0], ~\text{and} ~ [1 \, \alpha^i] ~ \text{for} ~ 0\le i\le q-2,
\label{eq:plp}
\end{equation}
where $\alpha$ is a primitive element of  $\mathbb{F}_q$.
Clearly, any network with two sources and arbitrary number of receives can be securely coded by reducing the set of available
colors in (\ref{eq:plp}) by removing point (color) $[1\, 1]$ and applying a wiretap code based on the matrix $H=[1\, 1]$ as in the
example above.
Alphabet size sufficient to securely code all network with two sources also follows from \cite{ceitnc}:
\begin{theorem}
\label{th:h2st2}
For any configuration with two sources $t$ receivers,
the code alphabet $\mathbb{F}_q$ of size
\[
\lfloor \sqrt{2t-7/4}+1/2 \rfloor +1
\label{eq_alph}
\]
is sufficient for a secure network code. There exist configurations for which it is necessary.
\end{theorem}

\begin{figure}[hbt]
\begin{center}
\psset{unit=0.045in}
\begin{pspicture}(0,0)(70,35)
\psset{linewidth=0.9pt}
\begin{small}
\cnode(35,32){1}{S}
\cnode[linecolor=white](28,38){1}{s1}
\cnode[linecolor=white](42,38){1}{sh}
\rput(35,29.5){$\cdots$}\rput(35,20){$\cdots$}\rput(35,5){$\cdots$}
\cnode(15,20){1}{u1}
\cnode(20,20){1}{u2}
\cnode[linecolor=white](30,20){1}{u3}
\cnode[linecolor=white](40,20){1}{u3r}
\rput(17,12){$\cdots$}\rput(53,12){$\cdots$}
\psarc[linecolor=blue]{->}(10,5){4}{10}{100}
\psarc[linecolor=blue]{<-}(60,5){4}{80}{170}
\psarc[linecolor=blue]{->}(35,32){4}{190}{350}
\rput(8.2,9){{$n$}}\rput(40,32.5){{$M$}}
\rput(61.8,9){{$n$}}
\ncline{->}{S}{u1}\ncline{->}{S}{u2}
\cnode(50,20){1}{u1r}
\cnode(55,20){1}{u2r}
\ncline{->}{S}{u1r}\ncline{->}{S}{u2r}
\cnode(10,5){1}{r1}
\ncline{->}{u1}{r1}\ncline{->}{u2}{r1}
\ncline{->}{u3}{r1}
\cnode(60,5){1}{r1r}
\ncline{->}{u1r}{r1r}\ncline{->}{u2r}{r1r}
\ncline{->}{u3r}{r1r}
\rput(10,2){{$R_1$}}
\rput(60,2){{$R_{\binom{M}{h}}$}} \end{small}
\end{pspicture}
\end{center} 
\caption{\label{fig:sn}Combination $B(n,M)$ network.}
\end{figure}
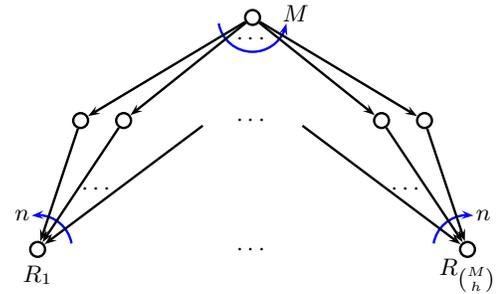

The wiretap approach to network security also provides the exact
alphabet size and secure code for a class of networks known as
combination networks and are illustrated in Fig.~\ref{fig:sn}.
There are $\binom{M}{n}$ receiver nodes. Note that each $n$ nodes of the second layer
are observed by a receiver. It is easy to see that an $[M+k,n]$ Reed Solomon code
can be used, namely, the first $k$ rows its parity check matrix can be used for the cosset code
and the rest as the coding vectors of the $M$ edges going out of the source.

\section{Connections with Other Schemes\label{sec:con}}

A number of connections between secure network coding with the concurrent work on network error correction can be observed
\cite{zhang06,yy07,matsumoto06}. We here describe the relationship between the proposed scheme and previously known constructions.
Cai and Yeung were first to study the design of secure
network codes for multicast demands \cite{yeung02secure}. They showed
that, in the setting described above, a secure network code can be
found for any $k\leq n-\mu$. Their construction is
equivalent to the following scheme:
\begin{enumerate}
\item Generate a vector $R=(r_1,r_2,\dots,r_{\mu})^T$ choosing its components uniformly
  at random over $\mathbb{F}_q$,
  \item Form vector $X$ by concatenating the $\mu$ random symbols $R$ to the $k$ source symbols $S$:
  \[
  X=\chsfk{S}{R}=(s_1,\dots,s_k,r_1,\dots,r_{\mu})^T
  \]
  \item Chose an \emph{invertible } $n\times n$ matrix over $\mathbb{F}_q$ and a linear code multicast (LCM) \cite{Li}
  to ensure the security condition (\ref{eqsecure}).
  (It is shown in \cite[Thm.~1]{yeung02secure} that such LCM and $T$ can be found provided that $q>\binom{|E|}{\mu}$.)
  \item Compute $Y=TX$ and multicast $Y$ to all the destinations by using the constructed code.
\end{enumerate}

Feldman \emph{et al.}\ considered also the same problem in \cite{feldman04csnc}.
Adopting the same approach of \cite{yeung02secure}, they showed that in order
for the code to be secure, the matrix $T$ should  satisfy certain
conditions (\cite[Thm.~6]{feldman04csnc}), that we restate here for convenience:
In the above transmission scheme, the security condition (\ref{eqsecure}) holds if and only if
any set of vectors consisting of
\begin{enumerate}
  \item at most  $\mu$ linearly independent global edge coding vectors and/or
  \item any number of vectors from the first $k$ rows of $T^{-1}$
\end{enumerate}
is linearly independent.
They also
showed that if one sacrifices in the number of information
packets, that is, take $k< n-\mu$, then one can find secure network
codes over fields of size much smaller than the very large bound $q>\binom{|E|}{\mu}$.

We will now show that our approach based on coding for the wiretap channel at the source is equivalent to
the above stated scheme \cite{yeung02secure} with the conditions of \cite{feldman04csnc}.
\begin{claim}
Let $T$ and ${\cal C}$ be a matrix and a corresponding secure
network code satisfying the above conditions. Set $H=T^*$ where
$T^*$ is the $k\times n$ matrix formed by taking the first $k$ rows
of $T^{-1}$. Then $H$ and ${\cal C}$ satisfy the condition of
Thm.~\ref{th:oursec}. \label{eqsecure}\end{claim}
\begin{proof}
Consider the secure multicast scheme of \cite{yeung02secure} as
presented above. For a given information vector $S\in
\mathbb{F}_q^k$, let $B(S)$ be the set of all possible vectors $Y\in
\mathbb{F}_q^{n}$ that could be multicast through the network under
this scheme. More precisely,
\begin{equation*}
    B(S)=\Bigl\{Y\in \mathbb{F}_q^n | Y=TX,  X=\chsfk{S}{R}, R\in\mathbb{F}_q^{n-k}\Bigr\}.
\end{equation*}
Then, for all $Y\in B(S)$, we have $ T^*Y=T^*T\chsfk{S}{T}=S. $
Therefore, any $Y\in B(S)$ also belongs to the coset of the space
spanned by the rows of $T^*$ whose syndrome is equal to $S$.
Moreover, since $T$ is invertible, $|B(S)|=2^{n-k}$ implying that
set $B(S)$ is exactly that coset. The conditions of
\cite{feldman04csnc} as stated above directly translate into
(\ref{eq:secc}), the remaining condition of Thm.~\ref{th:oursec}.
\end{proof}

\section{Conclusion}
We considered the problem of securing a multicast network
implementing network coding against a wiretapper capable of
observing a limited number of links of his choice, as defined initially by Cai and Yeung.
We showed that the problem can be formulated as a generalization of the wiretap channel of type II
(which was introduced and studied by Ozarow and Wyner), and
decomposed into two sub-problems: the first one of designing a secure wiretap channel code and the second of
designing a network code satisfying some additional constraints. We proved there is no penalty
to pay by adopting this separation, which we find in many ways illuminative.

\section*{Acknowledgments}
The authors would like to thank A.\ Sprintson for useful discussions about
this work and C.\ N.\ Georghiades for his continued support.
\bibliographystyle{ieeetran}{
\bibliography{ncsec}}
\end{document}